# Ionization dynamics and damage conditions in fused silica irradiated with Mid-Infrared femtosecond pulses

George D Tsibidis,[1, 2,a)] and Emmanuel Stratakis[1, 3,b)]

[1]*Institute of Electronic Structure and Laser (IESL), Foundation for Research and Technology (FORTH), Vassilika Vouton, 70013, Heraklion, Crete, Greece*

[2]*Department of Materials Science and Technology, University of Crete, 71003, Heraklion, Greece*

[3]*Department of Physics, University of Crete, 71003, Heraklion, Greece*

[a,b)] Authors to whom correspondence should be addressed: tsibidis@iesl.forth.gr; stratak@iesl.forth.gr

**ABSTRACT**

The employment of ultrashort laser sources at the mid-IR spectral region for dielectrics is expected to open innovative routes for laser patterning and a wealth of exciting applications in optics and photonics. To elucidate the material response to irradiation with mid-IR laser sources, a consistent analysis of the interaction of long wavelength femtosecond pulses with dielectric materials is presented. The influence of the pulse duration is particularly emphasized in specifying the laser parameters for which photoionization and impact ionization are important. Simulation results using pulses at 2.2 μm, 3.2 μm and 5 μm are conducted to illustrate the optimum conditions for the onset of damage on the solid that is related to the occurrence of the optical breakdown. Results predict that the damage threshold scales as $\sim\tau_p^a$ ($0.31 \leq a \leq 0.37$) at all laser wavelengths. Given the significant effect of the induced excitation level on the excitation of Surface Plasmons (SP) which account for the formation of laser-induced periodic structures (LIPSS) oriented perpendicular to the laser polarization, a correlation of the produced electron densities with SP and the threshold of SP excitation ($\sim\tau_p^\beta$, $0.33 \leq \beta \leq 0.39$) are also discussed in this as yet unexplored spectral region. Results are expected to guide development of an innovative approach to surface patterning using strong mid-IR pulses for advanced applications.

  The employment of femtosecond (fs) pulsed laser sources for material processing represents a precise technique with an impressive potential to fabricate a plethora of topographies for various technological applications [1-4]. In particular, the enhanced pattern complexity that can be produced is tailored in an unequivocal fashion through a modulation of the laser parameters (such as the energy, polarization states, energy dose, sequence of pulses, etc.). While a thorough investigation of laser-based patterning and exploration of the fundamental physical processes that characterize laser-matter interaction at wavelengths $\lambda_L < 1026$ nm has been exhaustively performed over the past decades, irradiation of solids with fs pulses and laser processing in the mid-Infrared (mid-IR) spectral region is a yet predominantly unexplored field despite some recent studies [5-16].

  One material that plays a very significant role in the design of optics in ultrashort pulsed laser systems is fused silica. Therefore, a deep understanding of the optical response, excitation levels and damage related mechanisms on fused silica upon exposure to high power mid-IR pulses [15-17] is very important as it will enable precise texturing that can be useful in various applications [8, 18]. On the other hand, elucidation of the interaction of mid-IR pulses with dielectrics can, also, provide crucial information for the behaviour of optical elements during experiments with intense mid-IR femtosecond pulses. Currently, the state of the art in the investigation of how mid-IR pulses interact with fused silica (FS) is not mature enough [14, 15, 19] to allow efficient development of innovative tools for mid-IR laser-based patterning and applications. More specifically, the absence of systematic efforts that allow evaluation of the impact of the combination of various laser parameters (i.e. fluence, pulse duration, wavelength) on the ultrafast dynamics, damage threshold and LIPSS formation at long wavelengths limits the optimization routes of the employment of mid-IR laser pulses for advanced applications. Thus, in this work, we investigate the ionization dynamics of FS upon excitation with mid-IR pulses and how the underlying effects could inflict material damage and surface patterning assuming a variety of laser conditions.

  To investigate the ultrafast dynamics and underlying physical phenomena, we, firstly, explore, the regimes in a range of spectral regions ($[\lambda_L<6$ μm]) at which the photoionization (PI) components, (i.e. Multiphoton Ionization (MPI) and/or Tunneling Ionization (TI) [20]) dominate; it is recalled that the PI mechanisms account for the formation of the free electron population ('seed electrons') in the conduction band (CB) which are important for the main ionization process, the avalanche (impact) ionization mechanism (AI) [21, 22]. It is known that the Keldysh parameter $\gamma \sim 1/(\sqrt{I}\lambda_L)$ (where $I$ stands for the laser intensity) is a crucial factor that defines where each PI component prevails (i.e. TI dominates at $\gamma \ll 1$ while MPI is the main contributor to PI at $\gamma \gg 1$) [20] and it can be used to determine the transition from MPI to TI. It is noted that an intermediate regime exists which is characterized by a coexistence of TI and MPI as a seed generation mechanism (region between $\gamma$=0.1 and $\gamma$=10 in Fig.1a). Simulation results illustrate a strong wavelength dependence associated with MPI which is reflected on the formation of ridges (Fig.1a,b). Results manifest that these ridges are developed for all wavelengths considered in this study. Each of the pronounced ridges (see also in Supplementary Material) corresponds to a range of photon energies and intensities for a particular order of MPI to occur. By contrast, the strong wavelength dependence of PI vanishes at higher intensities as the impact of MPI becomes less significant (Fig.1b).

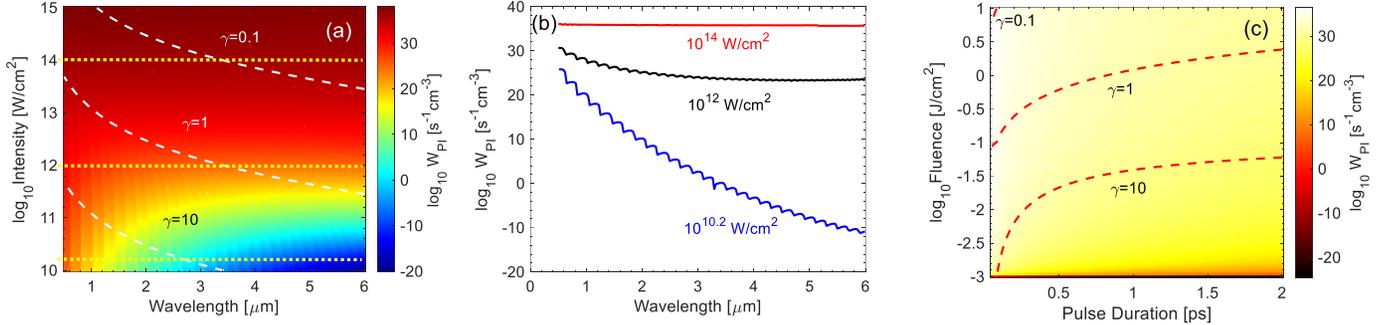

**FIG. 1.** Photoionization rates $W_{PI}$ as a function of (a) laser wavelength and (peak) intensity, (b) laser wavelength at three (peak) intensities, $10^{10.2}$ W/cm$^2$, $10^2$ W/cm$^2$ and $10^{14}$ W/cm$^2$ (*yellow* dotted lines in (a)), (c) laser (peak) fluence and pulse duration ($\lambda_L = 3.2$ μm).

In contrast to the traditional approach to determine the regimes at which MPI or/and TI become important (see [19] and references therein) through the laser intensity value, we follow an alternative approach in which we illustrate these regimes in a different representation where other laser parameters such as the (peak) fluence $F$ and pulse duration $\tau_p$ are involved. It is noted that throughout this work, a Gaussian temporal pulse shape for the beam has been used. On the other hand, any reference to the laser intensity ($\sim F/\tau_p$) hinders the individual impact of the two parameters. To evaluate the impact of the pulse duration and the laser fluence $F$ in PI and the contribution of TI and MPI, a detailed analysis is performed at $\lambda_L = 3.2$ μmThe results illustrate the dominant role of MPI at small fluences and a large range of pulse duration values while there is a combined effect of TI and MPI at moderately higher values of $F$. Finally, substantially high values of $F$ are required to lead to tunneling-assisted generation of the seed electrons in the conduction band (CB) (Fig.1c). Furthermore, considering the boundaries of the regions where MPI and TI dominate (defined by the *red* dashed lines in Fig.1c), results show that at smaller $\tau_p$ there is a gradual shift to smaller $F$ where the corresponding PI mechanisms are more pronounced. Similar conclusions can be deduced at other mid-IR wavelengths, however, an analysis of $\gamma$ indicates that there is a shift at which the aforementioned types of PI occur to lower pulse duration and higher (peak) fluences (see Section 3 in Supplementary Material).

Although the identification of the PI regions reveals which type of photoionization process becomes more significant in various laser conditions, a quantification of the excitation levels that are reached can be attained only through a detailed calculation of the electron dynamics. In the present work, the free electron generation in dielectrics following excitation with ultrashort pulsed lasers is described through the following two single rate equations which quantify both the evolution of the excited electron and self-trapped exciton (STE) densities, $N_e$ and $N_{STE}$ [19, 23], respectively,

$$\begin{aligned}\frac{dN_e}{dt} &= \frac{N_V - N_e}{N_V}\left(W_{PI}^{(1)} + N_e A^{(1)}\right) + \frac{N_{STE}}{N_V}\left(W_{PI}^{(2)} + N_e A^{(2)}\right) - \frac{N_e}{\tau_{tr}} \\ \frac{dN_{STE}}{dt} &= \frac{N_e}{\tau_{tr}} - \frac{N_{STE}}{N_V}\left(W_{PI}^{(2)} + N_e A^{(2)}\right)\end{aligned} \quad (1)$$

where $N_V$=2.2×10$^{22}$ cm$^{-3}$ corresponds to the valence electron density [23]. It is noted that other more rigorous approaches have also been developed which use a combination of (i) multiple rate equations (MRE) for the electron dynamics [22] and (ii) equations which accounts for the temporal change of the STE states [24, 25]. The MRE model was proposed to avoid the overestimation of the number of electrons participating in the avalanche procedure. Nevertheless, for the sake of simplicity, in this work, we aim to avoid the complexity derived from the employment of the more general model. Certainly, any approximation should not be made at the expense of a precise estimation of the strength of the underlying physical processes and therefore, the model should be validated through appropriate experimental protocols (see below).

In Eqs.1, the STE states are assumed to be centres situated at an energy level below the conduction band (i.e. $E_G^{(2)}$=6 eV) (see more details about STE in Section 6 in Supplementary Material); on the other hand, for fused silica, the band gap between the valence (VB) and the conduction band is $E_G^{(1)}$=9 eV [23]. Also, $\tau_r$~150 fs [26] stands for the trapping time of electrons in STE states which is usually longer than the laser pulse duration [19, 23] (see Supplementary Material). In the above framework, the excitation mechanism assumes photoionization ($W_{PI}^{(i)}$) and impact ionization processes (i.e. $A^{(i)}$ stands for the avalanche ionisation rate) that will lead transition from VB to CB ($i$=1) and the STE level to the CB ($i$=2); it is emphasized that, for the sake of simplicity, $W_{PI}^{(1)}$ is plotted in Fig.1 as the regions where the various components of photoionization occur are comparable if STE contribution is included. This behaviour has been reported in a previous work [19]. In Fig.2, $A^{(1)}$ is depicted as a function of the fluence and pulse duration for the transition from the valence band to CB for three wavelengths: 0.8 μm (to compare with results at lower laser wavelengths), 3.2 μm and 5 μm. Results indicate that at the same pulse duration and at increasing fluence, $A^{(1)}$ becomes larger at increasing wavelengths. Similarly, for the same fluence and at increasing pulse duration, $A^{(1)}$ becomes larger at increasing $\lambda_L$. This is due to the fact that $A^{(i)} \sim \lambda_L$ [19]. Similar results are obtained for the transition between STE level to the CB ($i=2$) (results are not shown). Thus, both levels lead to a similar avalanche ionization parameter despite a gap of ~3 eV (between the STE and CB) which means that the higher level does not influence the avalanche ionization rate. The increase of the impact ionization rate with the peak intensity (~$F/\tau_p$) or the electric field (i.e. the intensity is proportional to $cn\varepsilon_0|\vec{E}|^2$) have also been demonstrated in previous reports [27-29] (see also in Supporting Material). It is noted that results shown in Fig.2 illustrate the magnitude of the impact ionization rate at various combinations of fluence and pulse durations and they do not reflect the total excitation level; the total density of the excited carriers is derived from the accumulative effect of photoionization and impact ionization processes and it is discussed below (Eq.1).

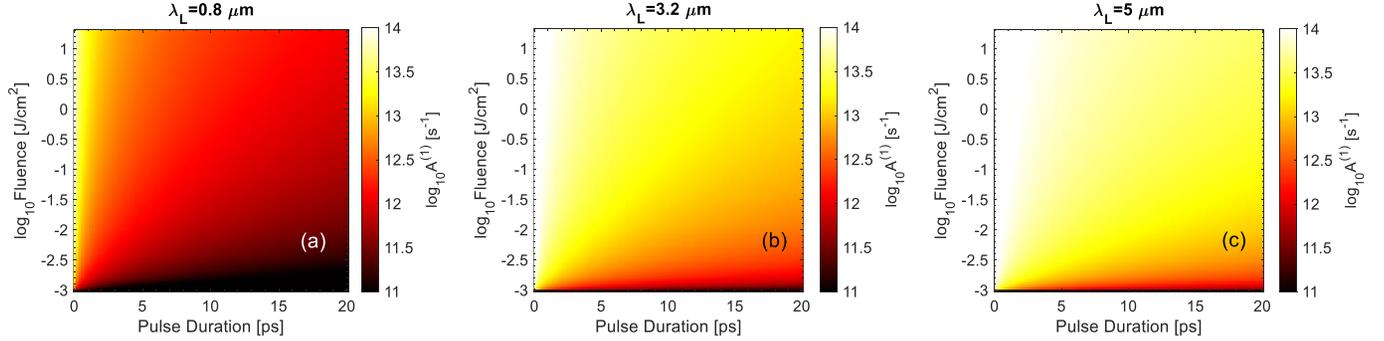

**FIG. 2.** Avalanche ionization rate at various laser (peak) fluences and pulse durations for (a) $\lambda_L = 0.8$ μm, (b) $\lambda_L = 3.2$ μm, and (c) $\lambda_L = 5.0$ μm.

The aforementioned set of equations is used to illustrate the dependence of the excited carrier densities $N_e$ on the laser conditions and to determine any specificity of irradiating with mid-IR pulses. To evaluate the role and influence of each ionization parameter, it is important to compare the impact of the photo- and avalanche- based ionization processes and significance of the presence of STE, separately. To this end, the predicted behaviour illustrated in Fig.3 provides a correlation of the excited electrons with the influence of the ionization processes assuming that the model includes contributions from: (a) both the impact ionization and photoionization processes and excitation of STE and (coined as 'Complete'), (b) the absence/presence of the STE states, (c) photoionisation-based processes (denoted by 'PI'), (d) combination of photoionisation and AI (coined as 'PI+AI').To compare the influence of the laser wavelength on the excitation levels reached, the theoretical model (Eq.1) was applied at short ($\lambda_L$=800 nm) and longer ($\lambda_L$=3200 nm) wavelengths. To allow a qualitative interpretation of the results, simulations were performed at the same laser (peak) intensity $I$=0.23×$10^{11}$ W/cm$^2$ at three different pulse durations ($\tau_p$=50 fs, 170 fs and 1 ps). Focusing, firstly, on results assuming $\lambda_L$=800 nm, it is evident that the impact ionisation contributes to the carrier excitation at all pulse durations. It is shown, though, that unlike a manifested pronounced role of the PI-based generation of excited carriers (see *blue solid* line) at short pulses there exists a significantly increased impact of AI (see *red dashed* line) at longer pulses which has also been reported in previous studies [27-29] (Fig.3a-c). Interestingly, a similar trend occurs at wavelengths in the mid-IR spectral region (Fig.3d-f); nevertheless, an enhanced influence of AI in the excited electron generation occurs at longer wavelengths which is attributed to the substantially larger avalanche ionisation rate (see Supplementary Material). Thus, at longer wavelengths, the generation of AI-assisted electrons yields a significantly larger contribution to the total density of the excited carriers which increases with pulse duration. In principle, the smaller excited carrier density $N_e$ due to PI which is expected at $\lambda_L$=3200 nm and reflects the decreasing $W_{PI}$ rate at longer wavelengths (Fig.1a) is outweighted by the larger number of electron densities produced due to AI. While simulation results obtained to emphasise the specificity of this particular wavelength are shown for $\lambda_L$=3200 nm, the analysis can be generalised for other values of $\lambda_L$.

A special emphasis is, also, drawn on the excitation of electrons form the STE levels assuming the above laser conditions. As explained above, the difference between the 'Complete' (depicted by the *black dotted* line in Fig.3) and the 'PI+AI' (see *red dashed* line) models is that the former includes the contribution of the STE states. According to the simulations, at very short pulses ($\tau_p$=50 fs), results derived from the employment of the 'Complete' and 'PI+AI' models are nearly identical which demonstrates that the contribution of the STE states in the generation of excited carriers is insignificant. This behaviour holds at both short (Fig.3a) and long wavelengths (Fig.3d). By contrast, as the pulse duration increases (while the peak intensity remains constant), electrons are derived that relax into the STE levels which subsequently are excited into the conduction band through PI- and AI-assisted processes. Hence, excitation of electrons that reside on STE through PI and AI appear to lead to an appreciable generation of carriers at long $\tau_p$. This trend indicates that a number of electrons in the STE states will not follow a relaxation and recombination process but will be re-excited through the two ionisation processes; this will result into a total $N_e$ *larger* than the electron density predicted for the case in which STE is not taken into account (see Supplementary Material). Although the STE electrons that will be re-excited were previously in the CB level upon the re-excitation process, they will contribute to the total $N_e$. It is noted that the 'PI+AI' model assumes that $N_{STE}(t) = 0$ and $\frac{dN_{STE}}{dt}(t) = 0$ (Eq. 1) at all times.

A parametric study has, also, been conducted to correlate the maximum carrier densities produced due to electron excitation with the values of the pulse duration and fluence for $\lambda_L = 2.2\ \mu m, 3.2\ \mu m$ and $5\ \mu m$ laser pulses. Simulation results allow the estimation of the fluence threshold $F_{thr}$ at which the (maximum) value of $N_e$ exceeds a critical value $N_e^{cr}$ (i.e. $N_e^{cr} \equiv 4\pi^2 c^2 m_e \varepsilon_0/(\lambda_L^2 e^2)$ that is coined as the optical breakdown threshold (OBT); in the expression that gives $N_e^{cr}$, $c$ is the speed of light, $m_e$ stands for mass of electron, $e$ is the electron charge and $\varepsilon_0$ is the permittivity of vacuum) while the calculation values for the OBT yield $N_e^{cr}$=2.31× 10$^{20}$ cm$^{-3}$ at $\lambda_L = 2.2$ μm, $N_e^{cr}$=1.09× 10$^{20}$ cm$^{-3}$ at $\lambda_L = 3.2$ μm, and $N_e^{cr}$=0.45× 10$^{20}$ cm$^{-3}$ at $\lambda_L = 5$ μm. These critical values are, usually, associated with the induced damage in dielectrics following exposure of the solid to intense heating [26] although thermal criteria (i.e. fluence values sufficiently high to induce material melting) were also used in other reports [19, 23]. According to our simulations (Fig.4a), $F_{thr}$ scales as $\sim\tau_p^a$ law ($a \cong 0.37$ for $\lambda_L = 2.2$ μm, $a \cong 0.34$ for $\lambda_L = 3.2$ μm, and $a \cong 0.31$ for $\lambda_L = 5$ μm). Previous experimental and theoretical studies have also demonstrated that similar power laws describe the damage threshold dependence on $\tau_p$ for various materials and laser wavelengths [14, 15, 28, 30-32]. Undoubtedly, appropriately designed experimental setups should be developed to validate the theoretical model and predictions. Apart from the critical plasma density related evaluation of damage threshold, other approaches have been used in previous reports to describe the onset of damage. In particular, at mid-IR, thermal criteria were used in semiconductors [30, 33] or metals [33] in which the theoretical model predictions exhibited a very good agreement with experimental results. A similar approach was also used to describe surface modification and LIPSS formation in fused silica [19]. Nevertheless, results reported in previous studies for the damage threshold following irradiation of fused silica at mid-IR [14, 15] or at lower wavelengths [15, 28] demonstrated an adequate agreement between the experimental observations and the selection of the critical density-based value as a criterion for the onset of material damage. In the present work, a comparison of the theoretical value from the model (Eq.1) with experimental data (Fig.4a) at similar laser conditions ($\lambda_L$~2 μm, $\tau_p$~150 fs) [15] demonstrates a satisfactory agreement without using thermal considerations. Despite the adequate estimation of $F_{thr}$ based on a critical density criterion, a more thorough future investigation assuming the contribution of thermal effects can potentially provide a more precise evaluation of the damage threshold. It is also important to evaluate the model assuming the role of the AI processes which is expected also to influence both the electron dynamics and the damage threshold. According to our simulations and the aforementioned discussion on the significant impact of AI excitation at long pulses and pulses at mid-IR, a plausible question can rise about whether the model is still valid in these conditions. Certainly, the agreement of the experimental data with the theoretical predictions for a wavelength in the mid-IR spectral region ($\lambda_L \sim 2$ μm) demonstrates the validity, however, more investigation is required, especially, at longer pulses; thus, the development of appropriate experimental protocols can allow the evaluation of the limitations of the model and requirement for further revision.

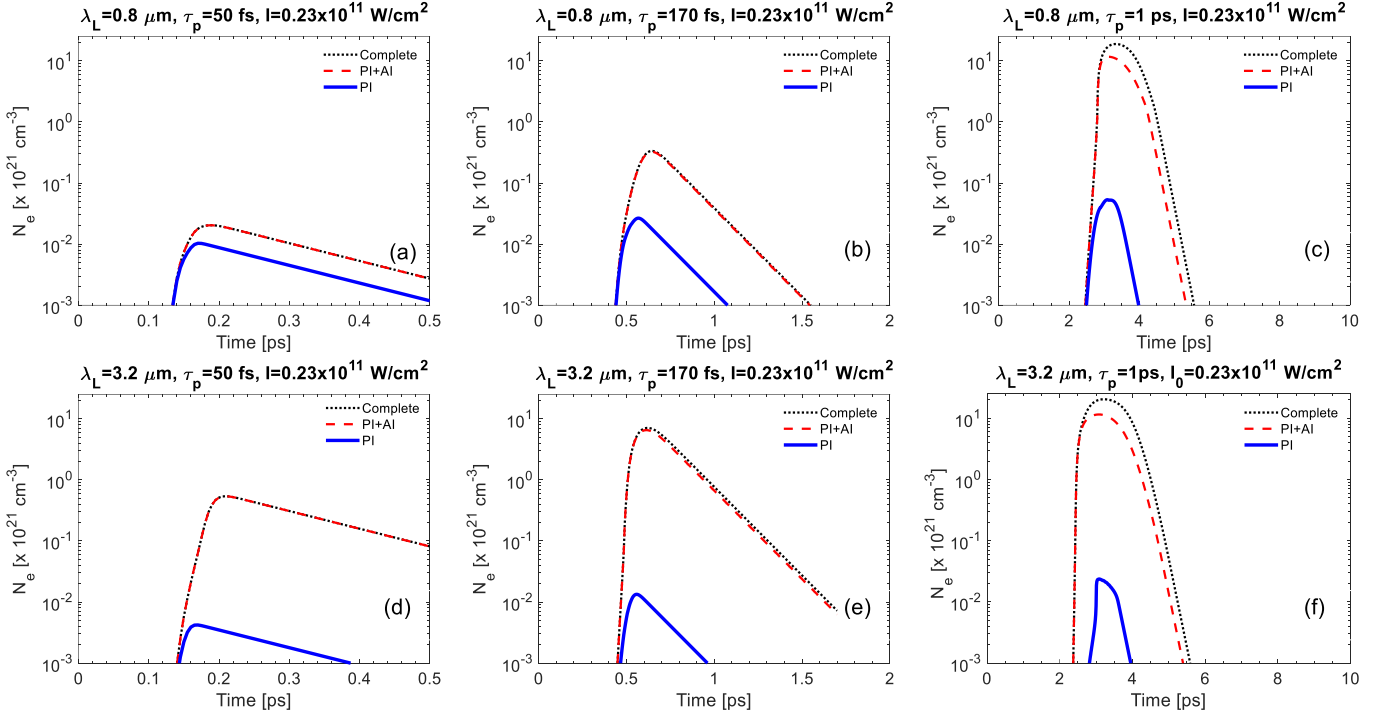

**FIG. 3.** Evolution of electron density assuming the complete model (*black dotted* line), PI+AI (*black dotted* line), PI for $\lambda_L$=0.8 μm ((a),(b),(c)) and $\lambda_L$=3.2 μm ((d), (e) and (f)) at three values of the pulse duration ($\tau_p$=50 fs, 170 fs and 1 ps) at the same peak laser intensity $I$=0.23×10$^{11}$ W/cm$^2$.

To avoid any confusion, it is also emphasised that the trend of decreasing damage threshold at increasing laser wavelength which has been in previous reports (for mid-IR [30] or at smaller wavelengths [27]) is not contradicted from results shown in Fig.4a; more specifically, results shown in Fig.4a do not illustrate the minimum fluence to reach the same value of temperature (melting point) at various wavelengths [30] but the minimum fluence to exceed $N_e^{cr}$ which is wavelength-dependent. Thus, higher values of $F_{thr}$ are expected to exceed larger $N_e^{cr}$ which increases at decreasing wavelength. Furthermore, interpretation of this behaviour has also been attributed to the impact of avalanche ionisation [27].

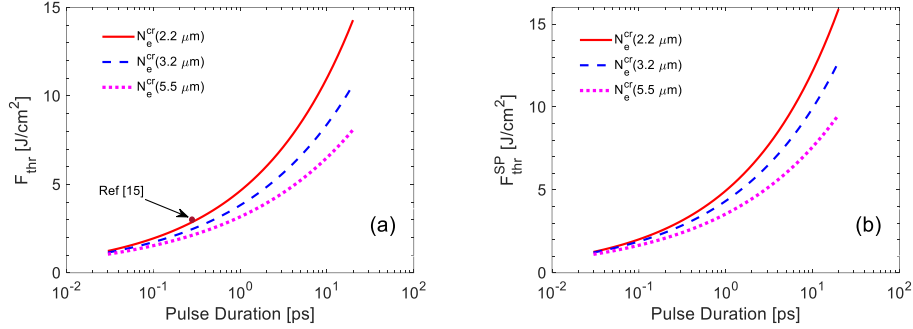

**FIG. 4.** (a) Minimum fluence to reach OBT as a function of laser pulse duration (for $\lambda_L = 2.2$ μm, 3.2 μm, 5 μm); *Brown* coloured filled circle corresponds to experimental data at 2.2 μm for 150 fs [15], (b) Threshold for SP excitation as a function of laser pulse duration (for $\lambda_L = 2.2$ μm, 3.2 μm, 5 μm).

Apart from the determination of the laser parameters required to induce material damage, further discussion can focus on how the aforementioned theoretical model is capable to assist in the evaluation of the conditions that allow surface texturing. Due to a large application potential, topographies covered with LIPSS constitute a very important class of surface patterns (see [1] and references therein). Several theoretical frameworks have been developed over the past years that aimed to elucidate the surface modification process and, more specifically, the formation of LIPSS. In principle, the predominant mechanism that accounts for the formation of self-assembled periodic structures [1, 34, 35] or even more complex structures [36] has been predominantly attributed to electromagnetic effects (i.e. excitation of scattered surface waves, including Surface plasmons (SP) [37, 38]). In particular, one class of LIPSS that are formed on FS, the Low Spatial Frequency LIPSS of sizes $\sim \lambda_L$ and oriented perpendicularly to the laser polarization (LSFL⊥) have been observed at lower wavelengths [32, 39]. In a previous report, we have systematically explored their formation at various wavelengths in the mid-IR spectral region [19] while their origin was associated with the excitation of Surface plasmon waves (see [1] and references therein). We recall that according to the SP-model, the calculated periodicity of the LSFL⊥ is provided by the expression $\Lambda = \lambda_L / Re\sqrt{\frac{\varepsilon}{\varepsilon+1}}$ [30, 37, 40] for irradiation in air which is approximately correct for a small number of pulses [41]. In the above expression, $\varepsilon$ stands for the carrier density dependent dielectric function of the irradiated material. According to the SP model, excitation of SP is possible if the carrier density exceeds a critical value that yields $Re(\varepsilon)<-1$ [30, 42]. Simulations for the threshold value at which the carrier density leads to excitation of SP provide the following results: $N_e^{SP} = 3.53 \times 10^{20}$ cm$^{-3}$ at $\lambda_L = 2.2$ μm, $N_e^{SP} = 1.68 \times 10^{20}$ cm$^{-3}$ at $\lambda_L = 3.2$ μm, $N_e^{SP} = 0.68 \times 10^{20}$ cm$^{-3}$ at $\lambda_L = 5$ μm. Comparing with $N_e^{cr}$, it appears that excitation of SP occurs at even higher values than those that lead to optical breakdown. The fact that $N_e^{cr} < N_e^{SP}$ (and not $N_e^{cr} > N_e^{SP}$) indicates that upon exceeding the OBT (i.e. $N_e^{cr} < N_e$), the laser conditions are capable to lead to excitation levels which can result to either (i) LSFL parallel (for $N_e^{cr} < N_e < N_e^{SP}$) to the laser polarization, LSFL∥, or (ii) LSFL⊥ (for $N_e > N_e^{SP}$). Both types of structures correspond to realistic scenarios (see Section 7 in Supplementary Material) [19,39]. Thus, the model can, also, be used to determine the conditions for the fabrication of LSFL⊥ or LSFL∥.

As one of the main aims of the current work is to determine the influence of mid-IR pulses on damage and texturing onset in various laser conditions, one plausible question is whether there is a correlation of the pulse duration with the minimum fluence value for SP excitation. To provide an estimate of the (minimum) fluence $F_{thr}^{SP}$ at which SP are excited, a thorough investigation was followed for a range of pulse durations and for three laser wavelengths. Results are illustrated in Fig.4b and it is shown that $F_{thr}^{SP}$ scales as $\sim \tau_p^\beta$ law ($\beta \cong 0.39$ for $\lambda_L = 2.2$ μm, $\beta \cong 0.36$ for $\lambda_L = 3.2$ μm, and $\beta \cong 0.33$ for $\lambda_L = 5$ μm).

As emphasized above, a validation of the model through the development of appropriate experimental protocols is required to both test the theoretical framework and define the laser conditions at which a revised version of the model should be introduced. For example, an overestimation of the AI-processes through the employment of a mid-IR-adapted multi-rate equations model can be reduced as it was performed at lower wavelengths [22, 25, 43]. The availability of experimental data for various laser conditions can also allow to define regimes in which the current model adequately describes the electron dynamics or more complex physical models are required. It is, also, very interesting to explore in more detail the ultrafast dynamics at very short pulses (<100 fs). In previous reports, it was shown that excitation of metals at very short pulses affects electron scattering and electron relaxation processes [44-47]; thus, a key question is whether appropriate modifications to Eqs.1 for dielectrics need to be introduced or the inclusion of Density Functional Theories coupled with a revised version of those equations is required. These issues constitute some of the fundamental questions that need to be addressed in a future study.

On the other hand, one of the predominant reasons of investigating laser patterning in the mid-IR spectral region is the capacity to texture patterns that have potentially different features from those at lower laser wavelengths. A recent investigation in metals and semiconductors showed interesting characteristics of the patterns that demonstrates the specificity of surface modification with mid-IR pulses (see [33] and Supplementary Material). Thus, an extension of the analysis on surface modification and patterning on FS would require a parametric study and theoretical exploration of multiphysical processes (on various temporal scales) including the investigation of thermal response and surface modification processes, however, such an exploration is beyond the scope of the present study.

In summary, our approach aimed to present a thorough analysis of the impact of the ionization processes in dielectrics in various laser conditions, identify the contribution of STE levels, define the regimes at which particular excitation mechanisms dominate, evaluate the pulse duration dependence of the damage thresholds and determine the conditions for potential formation of LSFL⊥ structures through the inclusion of a power law that relates SP excitation and the pulse duration. More specifically, in regard to the electron dynamics, it was shown that the impact ionization does not only play a significant role at longer pulses but also at longer wavelengths. Furthermore, while at both short and long wavelengths, the influence of STE is minimal at short pulses, at longer pulses it becomes significant. Therefore, it is important to incorporate the formation and re-excitation of STE in a theoretical framework for a precise estimation of the dynamics and OBT. Finally, it was shown that the predicted threshold for SP excitation is higher than the damage threshold which is important for the determination of the conditions for the fabrication of LSFL⊥ or LSFL∥. The methodology and results presented in this work could also provide feedback for ultrafast spectroscopy with mid-IR and reveal the role of the pulse duration in these spectroscopy studies. The ability to control the ultrafast dynamics of an irradiated material can provide innovative routes for optimizing the outcome of nano-processing through the employment of mid-IR laser pulses and use of long wavelength pulses in a wide range of mid-IR spectroscopy related applications.

See the supplementary material for the wavelength dependence of the photoionisation, the dependence of avalanche ionization rate on the laser intensity, a short description of the specificity of surface modification and excitation with mid-IR pulses and the contribution of the STE states to the electron excitation.

This work was supported from the following research projects: BioCombs4Nanofibres (grant agreement No. 862016); HELLAS-CH project (MIS 5002735); COST Action TUMIEE.

## AUTHOR DECLARATIONS

### Conflict of interest

The authors declare no conflicts of interest.

### Author Contributions

**George D. Tsibidis:** conceptualization of the work; physical modelling, simulations and interpretation of the results; writing-original draft; writing-review and editing; **Emmanuel Stratakis:** conceptualization of the work; interpretation of the results; writing-review and editing.

## DATA AVAILABILITY

The data underlying the results presented in this paper are not publicly available at this time but may be obtained from the authors upon reasonable request.

# Supplementary Material for

# Ionization dynamics and damage conditions in fused silica irradiated with Mid-Infrared femtosecond pulses

George D Tsibidis,[1, 2,a)] and Emmanuel Stratakis[1, 3,b)]

[1]*Institute of Electronic Structure and Laser (IESL), Foundation for Research and Technology (FORTH), Vassilika Vouton, 70013, Heraklion, Crete, Greece*

[2]*Department of Materials Science and Technology, University of Crete, 71003, Heraklion, Greece*

[3]*Department of Physics, University of Crete, 71003, Heraklion, Greece*

[a,b)] Authors to whom correspondence should be addressed: tsibidis@iesl.forth.gr; stratak@iesl.forth.gr

## 1. Trapping time $\tau_r$

The value of the trapping time considered in this study is $\tau_r \sim 150$ fs [1] that is usually longer than the laser pulse duration [2, 3]. It is noted that this value is possible to influence the electron dynamics of the system as longer pulse durations are also considered in this study. Thus, in a future study, a more detailed analysis should be conducted aiming to evaluate electron dynamics at different values of $\tau_r$.

## 2. Wavelength dependence of photoionisation

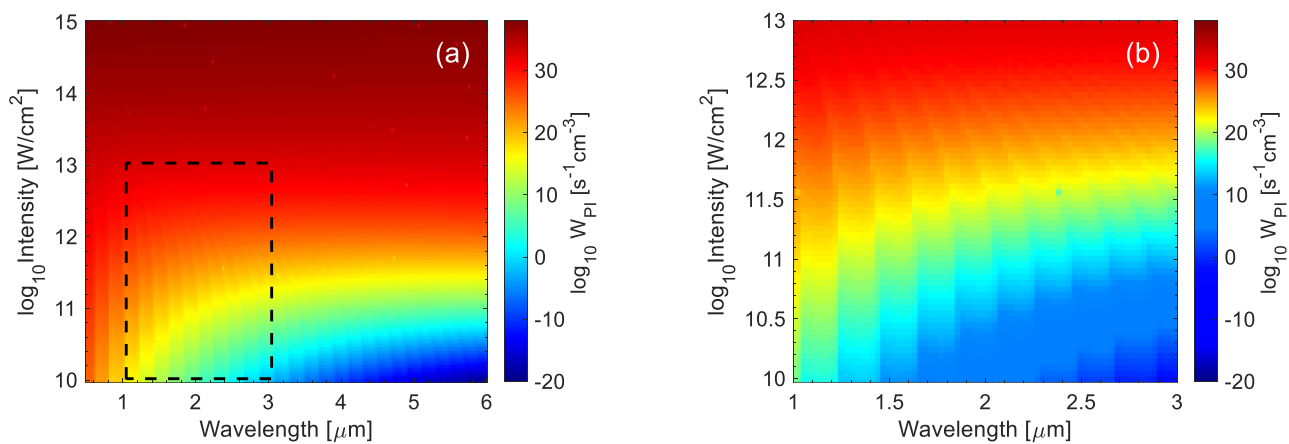

**FIG. 1S.** Photoionization rates $W_{PI}$ as a function of laser wavelength and (peak) intensity (a), (b) illustrates $W_{PI}$ inside the rectangular region in (a).

The photoionisation rate as a function of the wavelength and the intensity of the laser beam is illustrated in Fig.1Sa (it is the same as that in Fig.1a). Simulation results show a strong wavelength dependence associated with MPI which is reflected from the formation of ridges (Fig.1a,b). Results manifest that these ridges are developed for all wavelengths considered in this study. Each of the pronounced ridges corresponds to a range of photon energies and intensities for a particular order of MPI to occur. By contrast, the strong wavelength dependence of PI vanishes at higher intensities as the impact of MPI becomes less significant (Fig.1a, b and Fig.1Sa). To manifest the strong wavelength dependence of PI, the photoionisation rate inside the rectangular region in Fig.1Sa is shown in Fig.1Sb (in comparison with Fig.1Sa, the colormap has been modified to illustrate the profound ridges). To interpret the features of each ridge (at low intensities where the MPI character is more pronounced), we recall that the ionisation rate is proportional to $I^k$, where $I$ stands for the intensity and $k$ corresponds to the number of photons implied in the multiphoton ionisation. Thus, the ridge between 2.075 μm and 2.20 μm corresponds to a 16-photon absorption region while the next ridge (to the right of 2.2 μm) corresponds to a 17-photon absorption region.

## 3. Photoionisation rates as a function of fluence and pulse duration at different wavelengths

To show the impact of the laser conditions on the photoionization rate at different laser wavelengths, the photoionization rates $W_{PI}$ as a function of (peak) fluence and the pulse duration are illustrated in Fig.2S at $\lambda_L$= 0.8 nm, 3.2 nm and 5 nm. The Keldysh parameter $\gamma$ is also depicted to indicate the regions in which various photoionization mechanisms (multiphoton, tunneling or coexistence of multiphoton and tunneling) dominate (see main manuscript). In all figures, the colorbars have been scaled to the same range of values. Simulations show that an analysis of $\gamma$ indicates that there is a shift at which the aforementioned types of PI occur to lower pulse duration and higher (peak) fluences. On the other hand, a qualitative correlation of $W_{PI}$ as a function of (peak) fluence and the pulse duration at all three wavelengths looks similar.

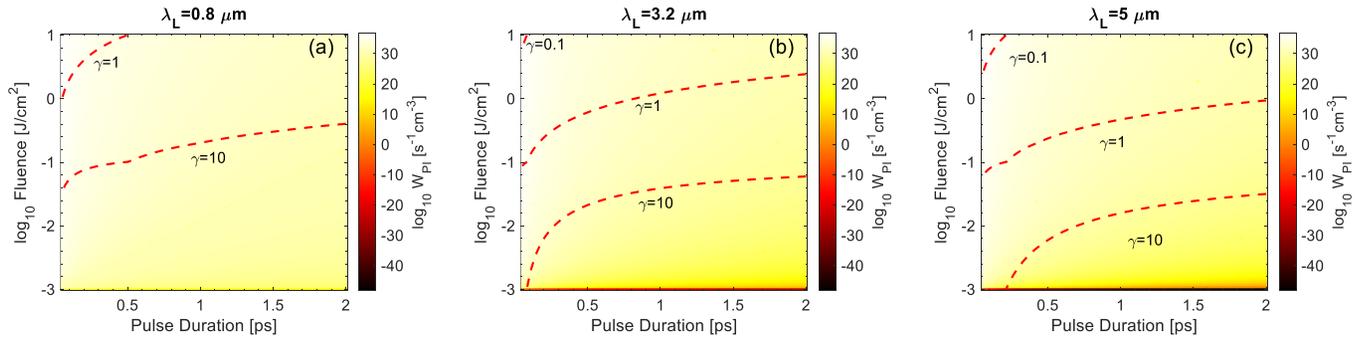

**FIG. 2S.** Photoionization rates $W_{PI}$ as a function of (peak) fluence and the pulse duration at (a) 0.8 nm, (b) 3.2 nm and (c) 5 nm.

### 4. Avalanche ionization rate dependence on laser intensity

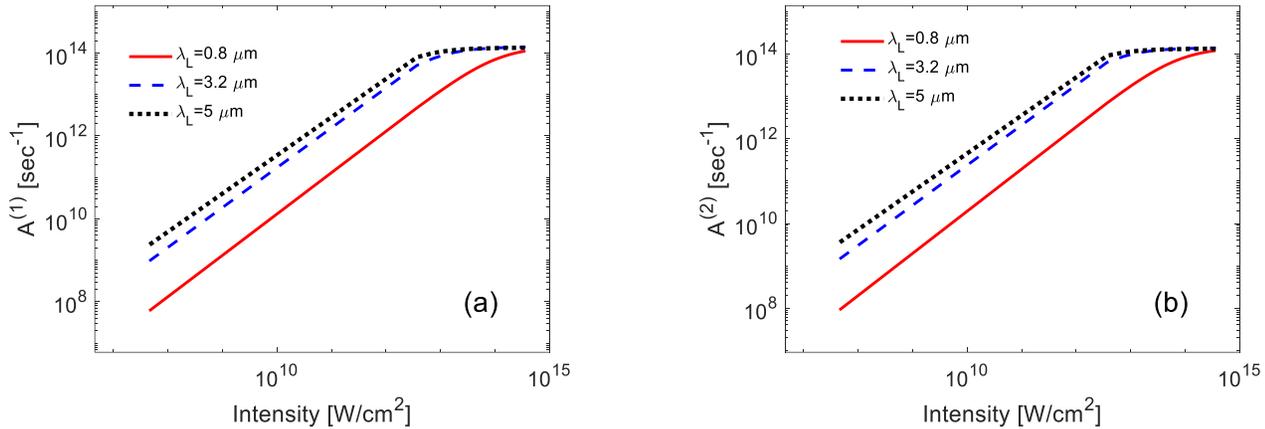

**FIG. 3S.** Avalanche ionisation rates (a) $A^{(1)}$ and (b) $A^{(2)}$ at various wavelengths.

$A^{(i)}$ ($i$=1,2) in Fig.3S stand for avalanche ionisation rate that will lead to transitions from Valence Band to Conduction Band ($i$=1) and the STE level to the Conduction Band ($i$=2) for three laser wavelengths.

### 5. Specificity of surface modification and excitation with mid-IR pulses

In previous reports, the thermal response of solids (Silicon [4, 5] and fused silica [2]) irradiated with fs mid-IR pulses was explored in detail in addition to the ultrafast dynamics. In particular, material modification was modelled in the latter case by using a multiphysical model on various temporal scales to describe relevant processes including energy absorption, electron excitation, relaxation, phase transitions and surface modification to derive the mechanism of surface modification and explain the formation of LIPSS. Although a direct comparison with results at lower wavelengths was not made, the different excitation levels, optical parameters such as the absorption coefficient (optical penetration depth) between irradiation at IR (or lower wavelengths) and mid-IR are expected to influence the thermal response of the material and therefore the features of the pattern (i.e. height, periodicity, etc) as the energy dose increases (at increasing number of pulses). Although the objective of the current work is to focus entirely on the role of excited carriers both on surface damage and conditions for onset of LSFL⊥ formation, we have also cited a recent work in which experimental results are presented to describe LIPSS formation following irradiation of silicon and nickel with mid-IR pulses [5]. A detailed parametric investigation at various fluences and number of pulses show that in contrast to LIPSS formed at lower laser wavelengths, interesting patterns are produced for silicon and nickel:

(i) For silicon: HSFL structures oriented parallel and perpendicular to the laser polarization (at much lower number of pulses than that for 800 nm, [6]), absence of suprawavelength structures (grooves) in contrast to their presence at lower wavelengths [7].

(ii)    For nickel: deep HSFL and HSFL structures are produced.

The above patterns indicate that there is a specificity in material modification at mid-IR.

Furthermore, results of excitation levels reached with mid-IR pulses show that at longer wavelengths (Fig.3 in the text), the generation of AI-assisted electrons yields a significantly larger contribution to the total density of the excited carriers which increases with pulse duration.

## 6. Contribution of STE states to electron excitation

The inclusion of the electron temperature in parameters such as the lifetime, relaxation time and plasma heating could provide interesting results to the electron excitation and relaxation. A coupling of a Two Temperature Model with Eq.1 (see [2,8]) could be used as an appropriate theoretical framework.

As noted in previous reports, STE are metastable states generated from the relaxation of free carrier relaxation and these defects correspond to *E* centers (i.e. oxygen vacancies) [9] situated ∼6 eV under the conduction band. They are, usually, produced in amorphous fused silica following ionization of the solid through laser irradiation which results into densification of the lattice [3,10]. A more detailed description on the conditions that lead to self-trapping is provided in [3,10].

## 7. Significance of $N_e^{SP} > N_e^{cr}$

As noted in the manuscript, SP excitation is related with the generation of LSFL⊥ structures (i.e. formation of LIPSS perpendicular to the laser polarisation with period of the size of $\lambda_L$). The fact that $N_e^{SP}$ is higher that $N_e^{cr}$ shows that for excitation levels (i.e. associated with a relevant produced electron density $N_e$) that lead to material damage (i.e. $N_e^{cr} < N_e$) the following two cases are possible:

(i)    for $N_e^{cr} < N_e < N_e^{SP}$ (due to the fact that the OBT has been exceeded) only LIPSS parallel to the laser polarisation will be produced as the condition for SP excitation is not satisfied.
(ii)   for $N_e > N_e^{SP}$ the onset of the formation of LSFL⊥ structures occurs as SP are excited.

Thus, if $N_e^{SP}$ was lower than $N_e^{cr}$, only LIPSS perpendicular to the laser polarisation will be formed upon exceeding the OBT which is not a realistic scenario for fused silica [11,12].